\documentclass[runningheads,envcountsame]{llncs}

\usepackage[utf8]{inputenc}
\usepackage{centernot}
\usepackage{amssymb}
\usepackage[normalem]{ulem}
\usepackage{centernot}
\usepackage{xspace}
\usepackage[bookmarks,unicode,colorlinks=true]{hyperref}
\usepackage{stmaryrd}
\usepackage{mathrsfs}
\usepackage{proof}
\usepackage{bm}
\usepackage{tikz}
\usepackage{multirow}
\usepackage{array}
\usepackage{mathtools}
\usepackage{enumitem}
\usepackage{cleveref}
\usepackage{microtype}
\usepackage{cite} 
\usepackage{adjustbox}
\usepackage{hhline}
\usepackage{pbox}

 \usepackage[location=inline]{moveproofs}
 \newcommand{\submission}[1]{}
 \newcommand{\report}[1]{#1}

\submission{
  \AtBeginDocument{%
    \addtolength\abovedisplayskip{-0.3\baselineskip}%
    \addtolength\belowdisplayskip{-0.3\baselineskip}%

    \setlist{nosep} 
    \setlist{itemsep=0pt, topsep=1pt}
  }

}

\makeatletter
\RequirePackage[bookmarks,unicode,colorlinks=true]{hyperref}%
   \def\@citecolor{blue}%
   \def\@urlcolor{blue}%
   \def\@linkcolor{blue}%

\def\orcidID#1{\smash{\href{http://orcid.org/#1}{\protect\raisebox{-1.25pt}{\protect\includegraphics{orcid_color.eps}}}}}
\makeatother

\hypersetup{
  pdftitle={Proving Non-Termination and Lower Runtime Bounds with LoAT},
  pdfauthor={Florian Frohn and Jürgen Giesl}
}

\renewcommand{\epsilon}{\varepsilon}
\let\oldphi\phi
\let\oldvarphi\varphi
\renewcommand{\phi}{\oldvarphi}
\renewcommand{\varphi}{\oldphi}
\renewcommand{\hat}[1]{\widehat{#1}}
\renewcommand{\check}[1]{\widecheck{#1}}

\newcommand{\TV}{\mathcal{TV}}
\newcommand{\LL}{\mathcal{L}}
\newcommand{\ZZ}{\mathbb{Z}}

\newcommand{\TT}{\mathcal{T}}

\newcommand{\NN}{\mathbb{N}}
\newcommand{\charfun}[1]{I_{#1}}
\newcommand{\Def}{\mathrel{\mathop:}=}
\newcommand{\assign}{\leftarrow}
\newcommand{\pl}[1]{\textsf{#1}}
\newcommand{\mDo}{\mathbf{do}}
\newcommand{\mWhile}[2]{\mathbf{while}\ #1\ \mDo\ #2}
\newcommand{\relmiddle}[1]{\mathrel{}\middle#1\mathrel{}}

\newcommand{\tool}[1]{\textsf{#1}}

\newcommand{\xto}[1]{\xrightarrow{#1}}
\newcommand{\cond}[1]{\left[#1\right]}
\newcommand{\prob}[4]{\left\llbracket #1 \relmiddle{|} #2 \relmiddle{|} #3 \right\rrbracket_{#4}}
\newcommand{\inc}{\mathit{inc}}
\newcommand{\dec}{\mathit{dec}}
\newcommand{\evdec}{{\mathit{ev}\text{-}\mathit{dec}}}
\newcommand{\evinc}{{\mathit{ev}\text{-}\mathit{inc}}}
\newcommand{\fixpoint}{\mathit{fp}}
\newcommand{\nt}{\mathit{nt}}
\newcommand{\AT}{AT}
\newcommand{\evaluationAlternatives}{\mathrel{\smash{\stackrel{\raisebox{3pt}{\scriptsize
          $k\:$}}{\smash{\to}}}_\TT^{m/*/+}}}
\newcommand{\evaluationTrans}{\mathrel{\smash{\stackrel{\raisebox{3pt}{\scriptsize
          $k\:$}}{\smash{\to}}}_\TT^*}}
\renewcommand{\emptyset}{\varnothing}

\newcommand{\nondet}{\mathit{nondet}}
\newcommand{\rc}{\mathit{rc}}
\newcommand{\dt}{\mathit{dh}}
\newcommand{\start}{\mathit{main}}
\newcommand{\sink}{\mathit{sink}}
\newcommand{\accel}{\mathit{accel}}
\newcommand{\encode}{\mathit{encode}}

\newcommand{\deps}{\mathit{deps}}
\newcommand{\closure}{\mathit{closure}}
\newcommand{\imp}{\mathit{imp}}
\newcommand{\vars}{\mathcal{V}\!\mathit{ars}}

\crefname{section}{Sect.}{Sect.}

\DeclareFontFamily{U}{mathx}{\hyphenchar\font45}
\DeclareFontShape{U}{mathx}{m}{n}{
      <5> <6> <7> <8> <9> <10>
      <10.95> <12> <14.4> <17.28> <20.74> <24.88>
      mathx10
      }{}
\DeclareSymbolFont{mathx}{U}{mathx}{m}{n}
\DeclareFontSubstitution{U}{mathx}{m}{n}
\DeclareMathAccent{\widecheck}{0}{mathx}{"71}

\crefname{equation}{eq.}{equations}%
\crefname{chapter}{chapter}{chapters}%
\crefname{section}{sect.}{sections}%
\crefname{appendix}{app.}{appendices}%
\crefname{enumi}{item}{items}%
\crefname{footnote}{footnote}{footnotes}%
\crefname{figure}{fig.}{figures}%
\crefname{table}{table}{tables}%
\crefname{theorem}{thm.}{theorems}%
\crefname{lemma}{lemma}{lemmas}%
\crefname{corollary}{cor.}{corollaries}%
\crefname{proposition}{proposition}{propositions}%
\crefname{definition}{def.}{definitions}%
\crefname{result}{result}{results}%
\crefname{example}{ex.}{examples}%
\crefname{remark}{remark}{remarks}%
\crefname{note}{note}{notes}%

 \report{
  \setlength{\textwidth}{125mm}
   \setlength{\textheight}{198mm}
 }

\title{Proving Non-Termination and Lower Runtime Bounds with \tool{LoAT}\report{\\} (System Description)\thanks{funded by
    the Deutsche Forschungsgemeinschaft (DFG, German Research Foundation)
    - 235950644 (Project GI 274/6-2).}}
\titlerunning{Proving Non-Termination and Lower Runtime Bounds with \tool{LoAT}}
\author{Florian Frohn\submission{\orcidID{0000-0003-0902-1994}} \and Jürgen Giesl\submission{\orcidID{0000-0003-0283-8520}}}
\institute{LuFG Informatik 2, RWTH Aachen University, Aachen, Germany}
\authorrunning{F.\ Frohn, J.\ Giesl}

\begin{document}

\maketitle

\begin{abstract}
  We present the\report{ new version of the} \emph{Loop Acceleration Tool} (\tool{LoAT}), a
powerful tool for proving non-termination and worst-case lower bounds for programs
operating on integers.
It is based on the novel calculus from \cite{tacas20,sttt21} for \emph{loop acceleration}, i.e., transforming loops into non-deterministic straight-line code, and for finding non-terminating configurations.
To implement it efficiently, \tool{LoAT} uses a new approach based on\report{ SMT solving and} unsat cores.%
\submission{
We evaluate \tool{LoAT}'s power and performance
by  extensive experiments.}%
\report{
An extensive evaluation shows that \tool{LoAT} is highly competitive with other state-of-the-art tools for proving non-termination.
While no other tool is able to deduce worst-case lower bounds for full integer programs, we also demonstrate that \tool{LoAT} significantly outperforms its predecessors from \cite{loat-journal,fmcad19}.
}

  \end{abstract}
\section{Introduction}
\label{sec:introduction}

Efficiency is one of the most important properties of software.
Consequently, \emph{automated complexity analysis} is of high interest to the software verification community.
Most research in this area has \report{(and still is) }focused on deducing \emph{upper} bounds on the worst-case complexity of programs\report{, which is useful to prove the absence of performance bugs}.
In contrast, the \emph{\underline{Lo}op \underline{A}cceleration \underline{T}ool} \tool{LoAT} aims to find performance bugs by deducing \emph{lower} bounds on the worst-case complexity of programs operating on integers.
\report{Besides finding performance bugs, a second motivation for lower complexity bounds is to complement upper bounds:
  If (asymptotically) identical lower and upper bounds can be deduced, then the (asymptotic) worst-case complexity of the program under consideration is known precisely.}%
\submission{Since non-termination implies the lower bound $\infty$, \tool{LoAT} is also equipped with non-termination techniques.}

\report{A particularly informative lower bound is $\infty$, meaning that there is no (finite) upper bound on the complexity of the program.
  This can happen for two reasons:
  Either the runtime of the program is always finite, but it depends on an unbounded non-deterministic assignment (as in ``$x \assign \nondet();\ \mWhile{x > 0}{x \assign x-1}$'') or the program does not terminate.
  \tool{LoAT} can detect (and distinguish) both sources of unbounded complexity.
  In particular, this means that it can also prove non-termination.
}

\tool{LoAT} is based on \emph{loop
acceleration}\cite{loat-journal,tacas20,sttt21,underapprox15,bozga10,bozga09a}, which
replaces loops by non-deterministic code:
The resulting program \report{non-deterministically }chooses a value $n$, representing the number of loop iterations in the original program.
To be sound, suitable constraints\submission{\linebreak} on $n$ are synthesized to ensure that the original loop allows for at least $n$ iterations\report{, given the current valuation of the program variables}.
Moreover, the transformed program updates the program variables to the same values as $n$ iterations of the original loop, but it does so in a single step.
To achieve that, the loop body is transformed into a \emph{closed form}, which is parameterized in $n$.
In this way, \tool{LoAT} is able to compute \emph{symbolic under-approximations} of programs, i.e., every execution path in the resulting transformed program corresponds to a path in the original program, but not necessarily vice versa.
In contrast to many other techniques for computing under-approximations\report{ (like
  \emph{bounded model checking} \cite{Clarke01})}, the symbolic approximations of \tool{LoAT} cover \emph{infinitely many runs} of \emph{arbitrary length}.
\report{Originally, the motivation for computing such approximations was that under-approximations that only regard runs of bounded length are usually unsuitable for deducing interesting (i.e., non-constant) lower complexity bounds.
  However, \tool{LoAT}'s symbolic under-approximations also turned out to be very valuable for use cases where bounded under-approximations suffice in theory, like proving \emph{reachability}:
  Together with suitable techniques for finding non-terminating configurations (see \Cref{sec:overview,sec:nonterm}), \tool{LoAT}'s ability to prove reachability turns it into an extremely powerful tool for proving non-termination (see \Cref{sec:experiments}).
}

\smallskip
\noindent
\emph{Contributions:}~ The main new feature of the novel version of \tool{LoAT} presented
in this paper is the \emph{integration} of the \emph{loop acceleration calculus} from
\cite{tacas20,sttt21}, which combines
different loop acceleration techniques in a modular way,
into \tool{LoAT}'s framework.
This enables \tool{LoAT} to use the loop acceleration calculus for the analysis of full integer programs, whereas the standalone implementation of the calculus from \cite{tacas20,sttt21} was only applicable to single loops without branching in the body.
To control the application of the
calculus, we use a new technique based on unsat cores (see \Cref{sec:sat}).%
\submission{ The new version of \tool{LoAT} is evaluated in extensive experiments. See \cite{report} for all proofs.}%
\report{ In particular, we present the following contributions:
  \begin{enumerate}
    \item The original implementation of the loop acceleration calculus in \cite{tacas20,sttt21} repeatedly tried to apply acceleration techniques to yet unprocessed inequations in the loop guard until all inequations were processed successfully, or no more progress was possible.
          In the worst case, such a strategy requires $\Theta(m^2)$ SMT queries to accelerate a loop whose guard consists of $m$ inequations.
          This can easily become a bottleneck in practice.
          In \Cref{sec:sat}, we introduce a novel technique which uses unsat cores to reduce the number of SMT queries to $\Theta(m)$.

    \item While the original implementation of the loop acceleration calculus in \cite{tacas20,sttt21} only considered single-path loops, we now integrated the calculus and our novel SMT-based approach to control its application into the framework of \tool{LoAT} (see \Cref{sec:overview}) to infer lower complexity bounds and non-termination proofs for full programs.
  \end{enumerate}
  Our evaluation in \Cref{sec:experiments} shows the benefits of the calculus over
  approaches based on monolithic loop acceleration techniques
  \cite{loat-journal,fmcad19}.
}

\section{Preliminaries}
\label{sec:prelim}

Let $\LL \supseteq \{\start\}$ be a finite set of \emph{locations}, where $\start$ is the \emph{canonical start location} (i.e., the entry point of the program), and let $\vec{x} \Def [x_1,\ldots,x_d]$ be the vector of \emph{program variables}.
Furthermore, let $\TV$ be a countably infinite set of \emph{temporary variables}, which
are used to model non-determinism, and let $\sup \ZZ \Def \infty$.
We call an arithmetic expression $e$ an \emph{integer expression} if it evaluates to an
integer when all variables in $e$ are instantiated by integers.
\tool{LoAT} analyzes tail-recursive programs operating on integers, represented as \emph{integer transition systems} (ITSs), i.e., sets of \emph{transitions} $f(\vec{x}) \xto{p} g(\vec{a}) \cond{\phi}$ where $f,g \in \LL$, the \emph{update} $\vec{a}$ is a vector of $d$ integer expressions over $\TV \cup \vec{x}$, the \emph{cost} $p$ is either an arithmetic expression over $\TV \cup \vec{x}$ or $\infty$, and the \emph{guard} $\phi$ is a conjunction of inequations over integer expressions with variables from $\TV \cup \vec{x}$.\footnote{
  \tool{LoAT} can also analyze the complexity of certain non-tail-recursive programs, see \cite{loat-journal}.
  For simplicity, we restrict ourselves to tail-recursive programs in the current paper.
}
For example, consider the loop on the left and the corresponding transition
\ref{eq:loop} on the right.

\vspace*{-.3cm}

\begin{minipage}{0.4\textwidth}
  \[
    \mWhile{x > 0}{x \assign x-1}
  \]
\end{minipage}
\begin{minipage}{0.45\textwidth}
  \begin{equation}
    \label{eq:loop}
    \tag{\(t_{\mathit{loop}}\)}
    f(x) \xto{1} f(x-1) \cond{x>0}
  \end{equation}
\end{minipage}

\medskip
\noindent
Here, the cost $1$ instructs \tool{LoAT} to use the number of loop iterations as cost measure.
\tool{LoAT} allows for arbitrary \emph{user defined} cost measures\report{ (like the number of arithmetic operations and comparisons)}, since the user can
choose any polynomials over the program variables as costs. 
\tool{LoAT} synthesizes transitions with cost $\infty$ to represent non-terminating runs, i.e., such transitions are not allowed in the input.

\report{ W.l.o.g., we assume that all inequations in $\phi$ are of the form $t > 0$ (as other inequations can be normalized accordingly, since all variables range over $\ZZ$), but we also use $\geq$ etc.\ in examples.}
A \emph{configuration} is of the form $f(\vec{c})$ with $f \in \LL$ and $\vec{c} \in \ZZ^d$.
For any entity $s \notin \LL$ and any arithmetic expressions $\vec{b} = [b_1,\ldots,b_d]$, let $s(\vec{b})$ denote the result of replacing each variable $x_i$ in $s$ by $b_i$, for all $1 \leq i \leq d$.
Moreover, $\vars(s)$ denotes the program variables and $\TV(s)$ denotes
the temporary variables occurring in $s$.
For an integer transition system $\TT$, a configuration $f(\vec{c})$ \emph{evaluates to
$g(\vec{c}\,')$ with cost $k \in \ZZ \cup \{\infty\}$}, written $f(\vec{c}) \xto{k}_\TT
g(\vec{c}\,')$, if
there exist a transition
$f(\vec{x}) \xto{p} g(\vec{a}) \cond{\phi} \in \TT$ and an instantiation of its temporary
variables with
integers such that the following holds:
\[
  \phi(\vec{c}) \land \vec{c}\,' = \vec{a}(\vec{c}) \land k = p(\vec{c}).
\]
\submission{ As usual, we write $f(\vec{c}) \evaluationTrans  g(\vec{c}\,')$ if $f(\vec{c})$
  evaluates to $g(\vec{c}\,')$ in arbitrarily many steps, and the sum of the costs of all steps is $k$.}%
\report{ As usual, we write $f(\vec{c}) \evaluationAlternatives g(\vec{c}\,')$ if $f(\vec{c})$ evaluates to $g(\vec{c}\,')$ in $m$ / arbitrary many / one or more steps, and the sum of the costs of all steps is $k$ (where $c + \infty = \infty + c = \infty$ for all $c \in \ZZ \cup \{\infty\}$).}
  We omit the costs if they are irrelevant.
The \emph{derivation height} of $f(\vec{c})$ is
\[
  \dt_\TT(f(\vec{c})) \Def \sup\{k \mid \exists g(\vec{c}\,'). \, f(\vec{c})
  \evaluationTrans g(\vec{c}\,')\} 
\] 
and the \emph{runtime complexity} of $\TT$ is 
\[
  \rc_\TT(n) \Def \sup\{\dt_\TT(\start(c_1,\ldots,c_d)) \mid |c_1| + \ldots + |c_d| \leq
  n\}. 
\]
$\TT$ terminates if no configuration $\start(\vec{c})$ admits an infinite ${\xto{}_\TT}$-sequence and $\TT$ is \emph{finitary} if no configuration $\start(\vec{c})$ admits a ${\xto{}_\TT}$-sequence with cost $\infty$.
Otherwise, $\vec{c}$ is a \emph{witness of non-termination} or a \emph{witness of infinitism}, respectively.
Note that termination implies finitism for ITSs where no transition
has cost $\infty$.
However, our approach may transform non-terminating ITSs into terminating, infinitary ITSs, as it replaces non-terminating loops by transitions with cost $\infty$.

\section{Overview of \tool{LoAT}}
\label{sec:overview}

The goal of \tool{LoAT} is to compute a lower bound on $\rc_\TT$ or even prove non-termination of $\TT$.
To this end, it repeatedly applies program simplifications, so-called \emph{processors}.
When applying them with a suitable strategy (see \cite{loat-journal,fmcad19}), one eventually obtains \emph{simplified transitions} of the form $\start(\vec{x}) \xto{p} f(\vec{a}) \cond{\phi}$ where $f \neq \start$.
As \tool{LoAT}'s processors are \emph{sound for lower bounds} (i.e., if they transform
$\TT$ to $\TT'$, then $\dt_\TT \geq \dt_{\TT'}$), such a simplified transition gives rise
to the lower bound $\charfun{\phi} \cdot p$ on $\dt_\TT(\start(\vec{x}))$ (where
$\charfun{\phi}$ denotes the indicator
function of $\phi$, which is $1$ for values where $\phi$ holds and $0$ otherwise).
\submission{This bound can be lifted to $\rc_\TT$ by solving a so-called \emph{limit problem}, see \cite{loat-journal}.}%
\report{%
  From these concrete bounds, \tool{LoAT} can deduce asymptotic lower bounds on $\rc_\TT$ by solving so-called \emph{limit problems}, which is achieved via an SMT encoding in most cases.
  For details on limit problems and techniques for solving them, we refer to \cite{loat-journal}.
}

\tool{LoAT}'s processors are also \emph{sound for non-termination}, as they preserve finitism.
So if $p = \infty$, then it suffices to prove satisfiability of $\phi$ to prove infinitism, which implies non-termination of the original ITS, where transitions with cost $\infty$ are forbidden (see \Cref{sec:prelim}).
\tool{LoAT}'s most important processors are:

\smallskip

\begin{description}
  \item[Loop Acceleration] (\Cref{sec:Modular Loop Acceleration}) transforms a
    \emph{simple loop}, i.e., a single transition  $f(\vec{x}) \xto{p}
    f(\vec{a}) \cond{\phi}$,
    into a non-de\-ter\-min\-istic transition that can simulate several loop iterations in one step.
    For example, loop acceleration transforms \ref{eq:loop} to
    \begin{equation}
      \label{eq:loop1}
      \tag{\(t_{\mathit{loop}^n}\)}
      f(x) \xto{n} f(x-n) \cond{x \geq n \land n > 0},
    \end{equation}
    where $n \in \TV$, i.e., the value of $n$ can be chosen non-deterministically.
  \item[Instantiation] \cite[Thm.\ 3.12]{loat-journal} replaces temporary variables by integer expressions.
    For example, it could instantiate $n$ with $x$ in \ref{eq:loop1}, resulting in\hspace*{0cm}
    \begin{equation}
      \label{eq:loop2}
      \tag{\(t_{\mathit{loop}^x}\)}
      f(x) \xto{x} f(0) \cond{x > 0}.
    \end{equation}
  \item[Chaining] \cite[Thm.\ 3.18]{loat-journal} combines two subsequent transitions into one transition.
    For example, chaining combines the transitions
    \begin{align}
      \report{\label{eq:start}
        \tag{\(t_{\mathit{init}}\)}}
      \submission{\notag}
      \phantom{\text{and \ref{eq:loop2} to}} \quad \start(x) &\xto{1} f(x)\\[-.1cm]
      \report{\tag{\(t_{\mathit{init}.\mathit{loop}^x}\)} \label{eq:chained}}
      \submission{\notag}
     \text{and \ref{eq:loop2} to} \quad
      \start(x) &\xto{x+1} f(0) \cond{x > 0}. \hspace*{2cm}
    \end{align} 
  \item[Nonterm] (\Cref{sec:nonterm}) searches for witnesses of non-termination, characterized by a formula $\psi$.
    So it turns, e.g.,
    \begin{align}
      \tag{\(t_{\mathit{nonterm}}\)}
      \label{eq:nonterm}
      \phantom{\text{into}} \quad f(x_1,x_2) &\xto{1} f(x_1-x_2,x_2) \cond{x_1 > 0} \\[-.1cm]
      \notag
      \text{into} \quad f(x_1,x_2) &\xto{\infty} \sink(x_1,x_2) \cond{x_1 > 0 \land x_2 \leq 0}
    \end{align}
    (where $\sink \in \LL$ is fresh), as each $\vec{c} \in \ZZ^2$ with $c_1 > 0 \land c_2
    \leq 0$ witnesses non-termination of \ref{eq:nonterm}, i.e., here $\psi$ is $x_1 > 0 \land x_2 \leq 0$.
\end{description}

\smallskip

Intuitively, \tool{LoAT} uses \textbf{Chaining}
to transform non-simple loops into simple loops. \pagebreak[3]
\textbf{Instantiation}  resolves non-determinism heuristically and thus reduces the number of temporary variables, which is crucial for scalability.
In addition to these processors, \tool{LoAT} removes transitions after processing them, as explained in \cite{loat-journal}.
\report{%
  To apply \textbf{Nonterm}, earlier versions of \tool{LoAT} tried to prove \emph{universal non-termination} (by proving $\phi \implies \phi(\vec{a})$, see \cite{loat-journal}), and they searched for \emph{recurrent sets} (via templates) and \emph{fixpoints}, i.e., recurrent singleton sets, (see \cite{fmcad19}).
  However, these techniques are no longer used.
  The reason is that the loop acceleration calculus from \cite{tacas20} can be adapted for detecting non-termination (see \cite{sttt21}), and using other specialized non-termination techniques in addition turned out to be superfluous.
}%
See \cite{loat-journal,fmcad19} for heuristics and a suitable strategy to apply \tool{LoAT}'s processors.

\report{
\begin{example}[Analyzing ITSs with {\normalfont{\tool{LoAT}}}]
  Consider the ITS consisting of the rules \ref{eq:start} and \ref{eq:loop}.
  \tool{LoAT} first accelerates \ref{eq:loop}, resulting in \ref{eq:loop1}.
  Then it instantiates \ref{eq:loop1}, resulting in \ref{eq:loop2}.
  Next, it chains \ref{eq:start} and \ref{eq:loop2}, resulting in \ref{eq:chained}.
  The resulting transition \ref{eq:chained} is simplified and gives rise to the concrete lower bound $I_{x>0} \cdot x$.
  By solving the limit problem that results from \ref{eq:chained}, \tool{LoAT} then deduces $\rc(n) \in \Omega(n)$.
\end{example}
}

\section{Modular Loop Acceleration}
\label{sec:Modular Loop Acceleration}

For \textbf{Loop Acceleration},
\tool{LoAT} uses \emph{conditional acceleration techniques} \cite{tacas20}. 
Given two formulas $\xi$ and $\check{\phi}$, and
a loop with update $\vec{a}$,
a conditional acceleration technique yields 
a formula $\accel(\xi, \check{\phi}, \vec{a})$
which implies that $\xi$ holds throughout $n$ loop iterations (i.e., $\xi$ is an
\emph{$n$-invariant}),   provided that $\check{\phi}$ is an $n$-invariant, too.
In the following,\submission{ \pagebreak[2]}  let $\vec{a}^0(\vec{x}) \Def \vec{x}$ and $\vec{a}^{m+1}(\vec{x}) \Def \vec{a}(\vec{a}^m(\vec{x})) = \vec{a}[\vec{x}/\vec{a}^m(\vec{x})]$.
\report{So in particular, $\vec{a}^1(\vec{x}) = \vec{a}$.}

\begin{definition}[Conditional Acceleration Technique]
  \label{def:acceleration technique}
A function $\accel$ is a \emph{conditional acceleration technique}
if the following implication holds for all formulas $\xi$ and $\check{\phi}$
with variables from $\TV \cup \vec{x}$,
all
updates $\vec{a}$,  all
$n>0$,
and all  instantiations of the variables with integers:
  \[
    \left(\accel(\xi, \check{\phi}, \vec{a}) \land \forall i \in [0,n).\ \check{\phi}(\vec{a}^i(\vec{x}))\right) \implies \forall i \in [0,n).\ \xi(\vec{a}^i(\vec{x})).
  \]
\end{definition}

\report{We only specify the result of conditional acceleration techniques for certain arguments and implicitly assume the result $\bot$\report{ (\emph{false})} for all other inputs.}
The prerequisite $\forall i \in [0,n).\ \check{\phi}(\vec{a}^i(\vec{x}))$ is ensured by
  previous acceleration steps, i.e., $\check{\phi}$ is initially $\top$ (\emph{true}),
  and it
  is refined by 
  conjoining a part $\xi$ of the loop guard in each acceleration step.
  \report{
    \begin{example}[Acceleration]
    \label{ex:conditional_acceleration}
    Reconsider the loop \eqref{eq:nonterm} with the update $\vec{a} = (x_1-x_2,x_2)$ and let $\check{\phi} = (x_2 \geq 0)$.
    Then we have
    \[
      (\forall i \in [0,n).\ \check{\phi}(\vec{a}^i(\vec{x}))) = (\forall i \in [0,n).\ x_2 \geq 0) \equiv x_2 \geq 0.
    \]
    Thus, for $\xi = (x_1>0)$ a conditional acceleration technique could yield
    \[
      \accel(x_1>0,\,x_2\geq 0, \, (x_1-x_2,x_2)) \; = \; (x_1 - (n-1) \cdot x_2 > 0),
    \]
    because if $x_2 \geq 0$, then $x_1 - (n-1) \cdot x_2 > 0$ implies $\forall i \in [0,n).\ x_1 - i \cdot x_2 > 0$ (i.e., it implies $\forall i \in [0,n).\ \xi(\vec{a}^i(\vec{x}))$), which would not be the case if $x_2 < 0$.
  \end{example}
  \noindent
  The novel version of \tool{LoAT} uses the following conditional acceleration techniques.
      }When formalizing acceleration techniques, we only specify the result of $\accel$ for
      certain arguments $\xi$, $\check{\phi}$, and $\vec{a}$,  and assume
 $\accel(\xi, \check{\phi}, \vec{a})  = \bot$  (\emph{false}) otherwise.   

  \begin{definition}[\tool{LoAT}'s Conditional Acceleration Techniques \cite{tacas20,sttt21}]
    \label{def:accel_techniques}\report{
    
 }   \resizebox{\textwidth}{!}{
      \begin{minipage}{\textwidth}
        \begin{tabbing}
          \textbf{Decrease} \= $\accel_\dec(\xi, \check{\phi}, \vec{a}) \Def \xi(\vec{a}^{n-1})$ \hspace*{.3cm} \=\kill \normalfont{\textbf{Increase}} \>$\accel_\inc(\xi, \check{\phi}, \vec{a}) \Def \xi$ \>if $\models \; \xi \land \check{\phi} \implies \xi(\vec{a})$\\
          \normalfont{\textbf{Decrease}}\>$\accel_\dec(\xi, \check{\phi}, \vec{a}) \Def \xi(\vec{a}^{n-1}(\vec{x}))$ \>if $\models \xi(\vec{a}) \land \check{\phi} \implies \xi$\\
          \normalfont{\textbf{Eventual Decrease}} \=$\accel_\evdec(t > 0, \check{\phi}, \vec{a}) \Def t > 0 \land t(\vec{a}^{n-1}(\vec{x})) > 0$ \\
          \>\>if $\models \; \left(t \geq t(\vec{a}) \land \check{\phi}\right) \implies t(\vec{a}) \geq t(\vec{a}^2(\vec{x}))$\\
          \normalfont{\textbf{Eventual Increase}}\>$\accel_\evinc(t > 0, \check{\phi},
          \vec{a}) \Def t > 0 \land t \leq t(\vec{a})$\\
          \>\>if $\models \; \left(t \leq t(\vec{a}) \land \check{\phi}\right) \implies t(\vec{a}) \leq t(\vec{a}^2(\vec{x}))$\\
          \normalfont{\textbf{Fixpoint}}\>$\accel_\fixpoint(t > 0, \check{\phi}, \vec{a}) \Def t>0 \land \bigwedge_{x \in \closure_{\vec{a}}(t)} x = x(\vec{a})$\\
          \>\>where $\closure_{\vec{a}}(t) \Def \bigcup_{i \in \NN} \vars(t(\vec{a}^i(\vec{x})))$
        \end{tabbing}
      \end{minipage}
    }
  \end{definition}

    The above five techniques are taken from \cite{tacas20,sttt21}, where only deterministic loops are considered (i.e., there are no temporary variables).
    Lifting them to non-deterministic loops in a way that allows for \emph{exact} conditional acceleration techniques (which capture all possible program runs) is non-trivial and beyond the scope of this paper.
    Thus, we sacrifice exactness and treat temporary variables like additional constant program variables whose update is the identity, resulting in a sound under-approximation (that captures a subset of all possible runs).
  \report{
    Then, the soundness proofs from \cite{tacas20,sttt21} trivially carry over to our setting, i.e., the techniques from \Cref{def:accel_techniques} are indeed conditional acceleration techniques as in \Cref{def:acceleration technique}.
    Note that in \cite{sttt21}, \textbf{Fixpoint} has been presented as a \emph{conditional non-termination technique}, a special case of \emph{conditional acceleration techniques}.
  }

  So essentially, {\bf Increase} and {\bf Decrease} handle inequations $t > 0$ in the loop
  guard where $t$ increases or decreases (weakly) monotonically when applying the loop's update.
\report{Then it suffices to require that $t > 0$ holds initially or right before the
  $n^{th}$ iteration, respectively. }%
The canonical examples where {\bf Increase} or {\bf Decrease} applies are
\[
  f(x,\ldots) \to f(x+1,\ldots) \cond{x>0 \land \ldots} \quad \text{or} \quad f(x,\ldots) \to f(x-1,\ldots) \cond{x>0 \land \ldots},
\]
respectively.
  {\bf Eventual Decrease} applies if $t$ never increases again once it starts to decrease.
\report{In such cases, it suffices to require that $t > 0$ holds initially \emph{and}
  right before the $n^{th}$ iteration. }%
The canonical example is
\submission{$f(x,y,\ldots) \to f(x+y,y-1,\ldots) \cond{x>0 \land \ldots}$.}
\report{\[f(x,y,\ldots) \to f(x+y,y-1,\ldots) \cond{x>0 \land \ldots}.\]}
\submission{Similarly, {\bf Eventual Increase} applies if $t$ never decreases again once it starts to increase.}%
\report{
  If $t$ never decreases again once it starts to increase, then {\bf Eventual Increase} applies and it suffices that $t$ increases and $t > 0$ holds.
  The canonical example is
  \begin{equation}
    \submission{\notag}
    \report{\tag{\text{ev-inc}}}
    \label{eq:evinc}
    f(x,y,\ldots) \to f(x+y,y+1,\ldots) \cond{x>0 \land \ldots}.
  \end{equation}
}
{\bf Fixpoint} can be used for inequations $t > 0$ that do not behave (eventually) monotonically.
\report{%
  In this case, $\accel_\fixpoint(t > 0, \check{\phi}, \vec{a})$ ensures that the value of $t$ remains constant when applying the body of the loop.
  The canonical example is
  \begin{equation}
    \submission{\notag}
    \report{\tag{\text{fixpoint}}}
    \label{eq:fixpoint}
    f(x, \ldots) \to f(-x, \ldots) \cond{x+1 > 0 \land \ldots}.
  \end{equation}
}%
It should only be used if $\accel_\fixpoint(t > 0, \check{\phi}, \vec{a})$ is satisfiable.
\report{%
  The first three conditional acceleration techniques are \emph{exact} for deterministic loops (i.e., loops that do not contain any temporary variables), see \cite{tacas20}.
  In contrast, this does not hold for {\bf Eventual Increase} and {\bf Fixpoint}.
  So if these techniques are used to accelerate a deterministic loop, then the resulting transition cannot necessarily simulate all runs of the original loop.
  For {\bf Eventual Increase}, the reason is that the sub-formula $t \leq t(\vec{a})$ requires that $t$ is always monotonically increasing, and hence the resulting transition does not cover runs where $t$ decreases initially.
  So for \eqref{eq:evinc}, {\bf Eventual Increase} yields the formula $x > 0 \land x \leq
  x + y \equiv x > 0 \land y \geq 0$, which does not cover the case that $y$ is negative initially.
  For {\bf Fixpoint}, the reason is that the resulting transition requires certain variables to remain constant.
  So for \eqref{eq:fixpoint}, it yields the formula $(x+1 > 0 \land x = -x) \equiv (x = 0)$.
}

\tool{LoAT} uses the \emph{acceleration calculus} of \cite{tacas20}.
It operates on \emph{acceleration problems}
$\prob{\psi}{\check{\phi}}{\hat{\phi}}{\vec{a}}$, where $\psi$ (which is initially
  $\top$) is repeatedly refined. When it stops, $\psi$ is used as the guard of the
  resulting accelerated transition. 
The formulas $\check{\phi}$ and $\hat{\phi}$ are the parts of the loop guard that have
already or have not yet been handled, respectively.
So $\check{\phi}$ is initially $\top$,
  and $\hat{\phi}$ and $\vec{a}$ are initialized with the guard $\phi$ and the
  update of the
  loop $f(\vec{x}) \xto{p} f(\vec{a}) \cond{\phi}$ under consideration, i.e., the
  initial acceleration problem is $\prob{\top}{\top}{\phi}{\vec{a}}$.
Once $\hat{\phi}$ is $\top$,\report{ one has reached a \emph{solved} acceleration problem $\prob{\psi}{\check{\phi}}{\top}{\vec{a}}$ and} the loop is accelerated to
\submission{$f(\vec{x}) \xto{q} f(\vec{a}^n(\vec{x})) \cond{\psi \land n > 0}$,}\report{\[f(\vec{x}) \xto{q} f(\vec{a}^n(\vec{x})) \cond{\psi \land n > 0},\]}
where the cost $q$ and a closed form for $\vec{a}^n(\vec{x})$ are computed by the recurrence solver \tool{PURRS} \cite{purrs}.
\report{Moreover, we assume that $\phi$ is satisfiable, as \tool{LoAT} does not try to accelerate a loop if satisfiability of its guard cannot be proven.}

\begin{definition}[Acceleration Calculus for Conjunctive Loops]
  \label{def:calculus}
  \report{%
    Let $\psi_1$, $\check{\phi}$, $\hat{\phi}$, and $\xi$ be formulas over $\TV \cup \vec{x}$ and let $\vec{a}$ be a vector of $d$ arithmetic expressions over $\TV \cup \vec{x}$.
  }%
  The relation ${\leadsto}$ on acceleration problems is defined as
  \[
    \infer[ \qquad
      \begin{array}{l}
        \accel \text { is a conditional} \\
        \text{acceleration technique}
      \end{array}
    ]{
      \prob{\psi_1}{\check{\phi}}{\xi \land \hat{\phi}}{\vec{a}} \leadsto \prob{\psi_1 \land \psi_2}{\check{\phi} \land \xi}{\hat{\phi}}{\vec{a}}
    }{
      \accel(\xi, \check{\phi}, \vec{a}) = \psi_2
    }
  \]
\end{definition}

So to accelerate a loop\report{ guard using ${\leadsto}$}, one picks a
not yet handled part $\xi$ of the guard in each step\report{ (typically a single inequation)}.
When  accelerating $f(\vec{x}) \xto{} f(\vec{a}) \cond{\xi}$ using a conditional
acceleration technique $\accel$, 
 one may assume $\forall i \in [0,n).\ \check{\phi}(\vec{a}^i(\vec{x}))$.
     The result of $\accel$ is conjoined
   to the\report{ partial} result $\psi_1$ computed so
far, and $\xi$ is moved from the third to the
second component of the\report{ acceleration} problem, i.e., to the already handled part
of the guard.\report{ This approach is \emph{modular}, since it handles every inequation of the guard separately and it can use different conditional acceleration techniques for different inequations in the same guard.}
\begin{example}[Acceleration Calculus]
  \label{eq:calculus}
  We show how to accelerate the loop
  \begin{align*}
    f(x,y) & \xto{x} f(x-y,y) \cond{x > 0 \land y \geq 0} \qquad \text{to}                                                                        \\
    f(x,y) & \xto{(x + \frac{y}{2}) \cdot n - \frac{y}{2} \cdot n^2} f(x-n \cdot y, y) \cond{y \geq 0 \land x - (n - 1) \cdot y > 0 \land n > 0}. \submission{\displaybreak[2]}
  \end{align*}
  The closed form $\vec{a}^n(x) = (x - n \cdot y, y)$ can be computed via recurrence solving.
  Similarly, the cost $(x + \frac{y}{2}) \cdot n - \frac{y}{2} \cdot n^2$ of $n$ loop
  iterations is obtained by solving the following recurrence relation
(where $c^{(n)}$ and $x^{(n)}$ denote the cost and the value of $x$ after $n$
  applications of the transition, respectively). 
  \[
    c^{(n)} = c^{(n-1)} + x^{(n-1)} = c^{(n-1)} + x - (n-1) \cdot y \qquad \text{and} \qquad c^{(1)} = x.
  \]
  The guard is computed as follows:
  \begin{align*}
    \prob{\top}{\top}{x > 0 \land y \geq 0}{\vec{a}} & \leadsto \prob{y \geq 0}{y \geq 0}{x > 0}{\vec{a}}                                           \\
                                                  & \leadsto \prob{y \geq 0 \land x - (n -
      1) \cdot y > 0}{y \geq 0 \land x > 0}{\top}{\vec{a}}.
  \end{align*}
  In the $1^{st}$ step, we have $\xi = (y \geq 0)$ and $\accel_\inc(y \geq 0, \top, \vec{a}) = (y \geq 0)$.
  In the $2^{nd}$ step, we have $\xi = (x > 0)$ and $\accel_\dec(x>0,y \geq 0,\vec{a}) = (x-(n-1) \cdot y > 0)$\report{ (as in \Cref{ex:conditional_acceleration})}.
  So the inequation $x - (n - 1) \cdot y > 0$ ensures $n$-invariance of
  $x>0$. 
\end{example}

\section{Efficient Loop Acceleration using Unsat Cores}
\label{sec:sat}

Each attempt to apply a conditional acceleration technique other than \textbf{Fixpoint} requires proving an implication\report{ (see \Cref{def:accel_techniques})}, which is implemented via SMT solving by proving unsatisfiability of its negation.
For \textbf{Fixpoint}, satisfiability of $\accel_\fixpoint(t > 0, \check{\phi}, \vec{a})$ is checked via SMT.
So even though \tool{LoAT} restricts $\xi$ to atoms, up to $\Theta(m^2)$ attempts to apply a
conditional acceleration technique are required to accelerate a loop whose guard contains
$m$ inequations using a naive strategy\submission{ ($5 \cdot m$ attempts for the $1^{st}$
  ${\leadsto}$-step, $5 \cdot (m-1)$ attempts for the $2^{nd}$ step, \ldots)}.
\report{%
  The reason is that the order of the inequations is not fixed and we cannot know in advance which conditional acceleration techniques apply to a given inequation.
  Hence, up to $5 \cdot m$ attempts are required for the $1^{st}$ ${\leadsto}$-step, up to $5 \cdot (m-1)$ are required for the $2^{nd}$ ${\leadsto}$-step, $\ldots$, resulting in up to $5 \cdot \frac{m^2 + m}{2}$ attempts in total.
  Thus, applying ${\leadsto}$ can easily become a bottleneck in practice.
}

To improve efficiency, \tool{LoAT} uses a novel encoding that requires just $5 \cdot m$ attempts.
For any $\alpha \in \AT_\imp = \{\inc, \dec, \evdec, \evinc\}$, let $\encode_\alpha(\xi, \check{\phi}, \vec{a})$ be the implication that has to be valid in order to apply $\accel_\alpha$, whose premise is of the form $\ldots \land \check{\phi}$.
Instead of repeatedly refining $\check{\phi}$, \tool{LoAT} tries to prove validity\footnote{Here and in the following, we unify conjunctions of atoms with sets of atoms.}
of $\encode_{\alpha,\xi} \Def \encode_\alpha( \xi, \phi \setminus \{\xi\}, \vec{a})$ for each $\alpha \in \AT_\imp$ and each $\xi \in \phi$, where $\phi$ is the (conjunctive) guard of the transition that should be accelerated.
Again, proving validity of an implication is equivalent to proving unsatisfiability of its negation.
So if validity of $\encode_{\alpha,\xi}$ can be shown, then SMT solvers can also provide an \emph{unsat core} for $\neg \encode_{\alpha,\xi}$.

\begin{definition}[Unsat Core]
  \label{Unsat Core}
  Given a conjunction $\psi$, we call each unsatisfiable subset of $\psi$ an \emph{unsat core} of $\psi$.
\end{definition}

\Cref{thm:core} shows that when handling an inequation $\xi$\report{ during acceleration, instead of the prerequisite that $\phi \setminus \{\xi\}$ is an $n$-invariant}, one only has to require $n$-invariance for the elements of $\phi \setminus \{\xi\}$ that occur in an unsat core of $\neg \encode_{\alpha,\xi}$.
Thus, an unsat core of $\neg \encode_{\alpha,\xi}$ can be used to determine which prerequisites $\check{\phi}$ are needed for the inequation $\xi$\report{ during acceleration}.
This information can then be used to find a suitable order for handling the inequations of the guard.
Thus, in this way one only has to check (un)satisfiability of the $4 \cdot m$ formulas $\neg \encode_{\alpha,\xi}$.
If no such order is found, then \tool{LoAT} either fails to accelerate the loop under consideration, or it resorts to using \textbf{Fixpoint}, as discussed below.

\begin{theorem}[Unsat Core Induces ${\leadsto}$-Step]
  \label{thm:core}
Let $\deps_{\alpha,\xi}$ be the intersection of $\phi \setminus \{\xi\}$ and an unsat core of $\neg \encode_{\alpha,\xi}$.  
  If $\check{\phi}$ implies $\deps_{\alpha,\xi}$, then $\accel_\alpha(\xi, \check{\phi}, \vec{a}) = \accel_\alpha(\xi, \phi \setminus \{\xi\}, \vec{a})$.
\end{theorem}
\report{
  \begin{proof}
    By \Cref{Unsat Core,def:accel_techniques}, unsatisfiability of $\neg \encode_{\alpha,\xi}$ implies unsatisfiability of $\neg \encode_\alpha(\xi, \check{\phi}, \vec{a})$.
    Thus, $\encode_\alpha(\xi, \check{\phi}, \vec{a})$ is valid, and hence the claim follows due to \Cref{def:accel_techniques}.
    \qed
  \end{proof}
}

\begin{example}[Controlling Acceleration Steps via Unsat Cores]
  \label{ex:encode}
  Reconsider \Cref{eq:calculus}.
  Here, \tool{LoAT} would try to prove\submission{, among others,} the following
  implications:
  \begin{align}
    \report{\encode_{\inc,x>0}   & \quad = \quad (x>0 \land y > 0) \implies x-y > 0                                        \\
    }
    \encode_{\dec,x>0}           & \quad = \quad (x-y>0 \land y > 0) \implies x > 0 \label{eq:dec_x} \\
    \report{\encode_{\evdec,x>0} & \quad = \quad (x \geq x-y \land y > 0) \implies x-y \geq x-2 \cdot y \label{eq:evdec_x}
    \\
    \encode_{\evinc,x>0}         & \quad = \quad (x \leq x-y \land y > 0) \implies x-y \leq x-2 \cdot y \label{eq:evinc_x}
    \\
    }
    \encode_{\inc,y>0}           & \quad = \quad (y>0 \land x > 0) \implies y > 0 \label{eq:inc_y}
    \report{                                                                                                               \\
    \encode_{\dec,y>0}           & \quad = \quad (y>0 \land x > 0) \implies y > 0 \label{eq:dec_y}
    \\
    \encode_{\evdec,y>0}         & \quad = \quad (y \geq y \land x > 0) \implies y \geq y \label{eq:evdec_y}
    \\
    \encode_{\evinc,y>0}         & \quad = \quad (y \leq y \land x > 0) \implies y \leq y \label{eq:evinc_y}
    }
  \end{align}
    To do so, it would try to prove unsatisfiability of $\neg \encode_{\alpha,\xi}$ via SMT.
  For \eqref{eq:dec_x}, we get $\neg \encode_{\dec,x>0} = (x-y > 0 \land y > 0 \land x
  \leq 0)$, whose only unsat core is $\neg \encode_{\dec,x>0}$,
\submission{\pagebreak[2]}
  and its intersection
   with $\phi \setminus \{x > 0\} = \{y > 0\}$ is $\{y > 0\}$.

  For \eqref{eq:inc_y}, we get $\neg \encode_{\inc,y>0} = (y > 0 \land x > 0 \land y \leq 0)$, whose minimal unsat core is $y > 0 \land y \leq 0$, and its intersection with $\phi \setminus \{y > 0\} = \{x > 0\}$ is empty.
  So by \Cref{thm:core}, we have $\accel_\inc(y>0, \top, \vec{a}) = \accel_\inc(y>0, x>0, \vec{a})$\report{ where $\vec{a} = (x-y,y)$}.

  \report{In this way, validity of \eqref{eq:dec_x} -- \eqref{eq:evinc_y} can be proven,
    and each proof gives rise to a corresponding unsat core.}
  \submission{In this way, validity of $\encode_{\alpha_1,x>0}$
and $\encode_{\alpha_2,y>0}$
    is proven for all $\alpha_1
    \in \AT_\imp \setminus \{ \inc\}$  and all $\alpha_2 \in \AT_\imp$.
    However,
the premise $x \leq x-y \land y > 0$ of $\encode_{\evinc,x>0}$ is unsatisfiable and thus a corresponding
acceleration step would yield a transition with unsatisfiable guard. To prevent that,
\tool{LoAT} only uses a technique $\alpha \in \AT_\imp$ for $\xi$ if
the premise of $\encode_{\alpha,\xi}$ is
satisfiable.
  }    
\end{example}

\report{
  Note that the premise of the implication \eqref{eq:evinc_x} in \Cref{ex:encode} is unsatisfiable.
  Consequently, a ${\leadsto}$-derivation that uses the acceleration technique \textbf{Eventual Increase} to deal with the inequation $x > 0$ leads to an accelerated transition with an unsatisfiable guard, even though the guard of the original loop is clearly satisfiable.
  This is possible since, as mentioned above, \textbf{Eventual Increase} is an under-approximating acceleration technique even if the loop under consideration is deterministic.
  Hence, the resulting formula may only cover a (possibly empty) subset of the runs that are admissible with the original loop.
  The same may happen for other conditional acceleration techniques from $\AT_\imp$ if they are applied to non-deterministic loops, where all of them may yield under-approximations due to our treatment of temporary variables as constants.
  To prevent that, \tool{LoAT} only uses such a conditional acceleration technique if satisfiability of the premise of the resulting implication can be proven, i.e., it would not consider the acceleration technique \textbf{Eventual Increase} for the inequation $x > 0$ in \Cref{ex:encode}.
}

So for each inequation $\xi$ from $\phi$, \tool{LoAT} synthesizes up to $4$ potential
${\leadsto}$-steps corresponding to $\accel_\alpha(\xi, \deps_{\alpha,\xi}, \vec{a})$,
where $\alpha \in \AT_\imp$.
If validity of $\encode_{\alpha,\xi}$ cannot be shown for any $\alpha \in \AT_\imp$, then \tool{LoAT} tries to prove satisfiability of $\accel_\fixpoint(\xi, \top, \vec{a})$ to see if \textbf{Fixpoint} should be applied.
Note that the $2^{nd}$ argument of $\accel_\fixpoint$ is irrelevant, i.e., \textbf{Fixpoint} does not benefit from previous acceleration steps and thus ${\leadsto}$-steps that use it do not have any dependencies.
\report{If satisfiability cannot be proven, then no acceleration technique is applied to $\xi$ and \tool{LoAT} gives up on accelerating the loop under consideration.}

It remains to find a suitably ordered subset $S$ of $m$ ${\leadsto}$-steps that constitutes a successful ${\leadsto}$-sequence\report{ and hence gives rise to the guard of the accelerated transition}.
\report{In particular, the dependencies among the ${\leadsto}$-steps must not be cyclic,
  i.e., the order on $S$ must be strict (i.e., irreflexive) and hence, well founded.
  The following lemma shows how to obtain the set $S$, together with a suitable order.
  Here,
}%
\submission{In the following, }%
we define $\AT \Def \AT_\imp \cup \{\fixpoint\}$ and we extend the definition of $\deps_{\alpha,\xi}$ to the case $\alpha = \fixpoint$ by defining $\deps_{\fixpoint,\xi} \Def \emptyset$.
\begin{lemma}
  \label{lem:order}
  Let $C \subseteq \AT \times \phi$ be the smallest set such that $(\alpha,\xi) \in C$ implies
  \begin{itemize}
    \item[(a)] if $\alpha \in \AT_\imp$, then $\encode_{\alpha,\xi}$ is valid and its premise is satisfiable,
    \item[(b)] if $\alpha = \fixpoint$, then $\accel_\fixpoint(\xi, \top, \vec{a})$ is satisfiable, and
    \item[(c)] $\deps_{\alpha,\xi} \subseteq \{\xi' \mid (\alpha',\xi') \in C \text{ for some } \alpha' \in \AT \}$.
  \end{itemize}
  Let $S \Def \{(\alpha,\xi) \in C \mid \alpha \geq_{\AT} \alpha' \text{ for all } (\alpha',\xi) \in C\}$ where ${>_{\AT}}$ is the total order $\inc >_{\AT} \dec >_{\AT} \evdec >_{\AT} \evinc >_{\AT} \fixpoint$.
  We define $(\alpha',\xi') \prec (\alpha,\xi)$ if $\xi' \in \deps_{\alpha,\xi}$.
  Then ${\prec}$ is a strict\submission{ (and hence, well-founded)} order on $S$.
\end{lemma}
\report{
  \begin{proof}
    Let $C'$ consist of all $(\alpha, \xi) \in \AT \times \phi$ which satisfy the conditions (a) and (b).
    Then due to the minimality of $C$, we have $C = C_0 \cup C_1 \cup \ldots$, where for all $i \in \NN$ we define
    \[
      \begin{array}{rcl}
        C_0     & = & \{ (\alpha, \xi) \in C' \mid \deps_{\alpha,\xi} = \emptyset \}                                                                    \\
        C_{i+1} & = & \{ (\alpha, \xi) \in C' \mid                                                                                                      \\
                &   & \qquad \deps_{\alpha,\xi}\subseteq \{\xi' \mid (\alpha',\xi') \in C_0 \cup \ldots \cup C_i \text{ for some } \alpha' \in \AT \}\}
        \setminus C_i.
      \end{array}
    \]
    Since $\AT \times \phi$ is finite, if $\prec$ were not well founded on $S$, then there would be $(\alpha_1,\xi_1), \ldots, (\alpha_n,\xi_n) \in \AT \times \phi$ such that
    \[
      (\alpha_1,\xi_1) \prec (\alpha_2,\xi_2) \prec \ldots \prec (\alpha_n,\xi_n) \prec (\alpha_1,\xi_1).
    \]
    Since $S \subseteq C$, for all $1 \leq j \leq n$, there exists an $i_j$ such that $(\alpha_j,\xi_j) \in C_{i_j}$.
    As $(\alpha',\xi') \prec (\alpha,\xi)$ implies $\xi' \in \deps_{\alpha,\xi}$, we have $i_1 < i_2 < \ldots < i_n < i_1$ which is a contradiction.
    \qed
  \end{proof}
}
The order ${>_{\AT}}$ in \Cref{lem:order} corresponds to the order \report{of acceleration techniques }proposed in \cite{tacas20}.
Note that the set $C$ can be computed without further (potentially expensive) SMT queries by a straightforward fixpoint iteration, and well-foundedness of ${\prec}$ follows from minimality of $C$.
For \Cref{ex:encode}, we get
\begin{align*}
  C = {} & \{(\dec, x>0), (\evdec, x>0)\} \cup \{(\alpha, y>0) \mid \alpha \in \AT\} & \text{and}        \\
  S = {} & \{(\dec, x>0), (\inc, y>0)\} \text{ with } (\inc, y>0) \; {\prec} \; (\dec, x>0).
\end{align*}
Finally, we can construct a valid ${\leadsto}$-sequence via the following theorem.
\begin{theorem}[Finding ${\leadsto}$-Sequences]
  \label{thm:order}
  Let $S$ be defined as in \Cref{lem:order} and assume that for each $\xi \in \phi$, there is an $\alpha \in \AT$ such that $(\alpha, \xi) \in S$.
  W.l.o.g., let $\phi = \bigwedge_{i=1}^m \xi_i$ where $(\alpha_1, \xi_1) \prec' \ldots \prec' (\alpha_m, \xi_{m})$ for some strict total order ${\prec'}$ containing ${\prec}$, and let $\check{\phi}_{j} \Def \bigwedge_{i=1}^j \xi_i$.
  Then for all $j \in [0,m)$, we have:
  \[
    \resizebox{\textwidth}{!}{
      $ \prob{\bigwedge_{i=1}^j \accel_{\alpha_i}(\xi_i, \check{\phi}_{i-1}, \vec{a})}{\check{\phi}_{j}}{\bigwedge_{i=j+1}^m \xi_i}{\vec{a}}
        \leadsto \prob{\bigwedge_{i=1}^{j+1} \accel_{\alpha_i}(\xi_i, \check{\phi}_{i-1}, \vec{a})}{\check{\phi}_{j+1}}{{\bigwedge_{i=j+2}^m \xi_i}}{\vec{a}}
      $
    }
  \]
\end{theorem}
\report{
  \begin{proof}
    First note that for all $j \in [0,m)$, we have $\deps_{\alpha_{j+1}, \xi_{j+1}} \subseteq \{ \xi_1, \ldots, \xi_j \}$.
    The reason is that $\xi_i \in \deps_{\alpha_{j+1}, \xi_{j+1}}$ implies $(\alpha_i,\xi_i) \prec (\alpha_{j+1}, \xi_{j+1})$ and hence, $(\alpha_i,\xi_i) \prec' (\alpha_{j+1}, \xi_{j+1})$, since $\prec'$ contains $\prec$.
    If $i > j$, then we would have $(\alpha_i,\xi_i) \prec' (\alpha_{j+1}, \xi_{j+1})\prec'(\alpha_i,\xi_i)$, which contradicts the irreflexivity of the strict order $\prec'$.

    So for all $j \in [0,m)$, if $\alpha \in \AT_\imp$ then $\bigwedge_{i=1}^j \xi_i$ implies $\deps_{\alpha_{j+1}, \xi_{j+1}}$, i.e., $\check{\phi}_{j}$ implies the intersection of an unsat core of $\neg \encode_{\alpha_{j+1}, \xi_{j+1}}$ and $\phi \setminus \{\xi_{j+1}\}$.
    Therefore, we have $\accel_{\alpha_{j+1}}(\xi_{j+1}, \check{\phi}_j, \vec{a}) = \accel_{\alpha_{j+1}}(\xi_{j+1}, \phi \setminus \{\xi_{j+1}\}, \vec{a})$ by \Cref{thm:core}.
    Thus, \Cref{thm:order} describes a correct $\leadsto$-step according to \Cref{def:calculus}.
    \qed
  \end{proof}
}

In our example, we have ${\prec'} = {\prec}$ as ${\prec}$ is total.
Thus, we obtain a ${\leadsto}$-sequence by first processing $y>0$ with \textbf{Increase} and then processing $x>0$ with \textbf{Decrease}.
\report{%
  In practice, for each $\xi \in \phi$, \tool{LoAT} synthesizes potential ${\leadsto}$-steps by considering $(\alpha, \xi)$ before $(\alpha', \xi)$ if $\alpha >_{\AT} \alpha'$, and it stops as soon as it finds an $\alpha$ where the preconditions $\deps_{\alpha,\xi}$ can already be satisfied without introducing cyclic dependencies, i.e., if $(\alpha,\xi)$ can be added to $S$ without contradicting well-foundedness of ${\prec}$.
  So in \Cref{ex:encode}, \tool{LoAT} would not check the implications \eqref{eq:dec_y} -- \eqref{eq:evinc_y}.
}

\section{Proving Non-Termination of Simple Loops}
\label{sec:nonterm}

To prove non-termination, \tool{LoAT} uses a variation of the calculus from \Cref{sec:Modular Loop Acceleration}, see \cite{sttt21}.
To adapt it for proving non-termination, further restrictions have to be imposed on the
conditional acceleration techniques, \submission{\pagebreak[3]}%
resulting in the notion of \emph{conditional non-termination techniques}, see \cite[Def.\ 10]{sttt21}.
We denote a ${\leadsto}$-step that uses a conditional non-termination technique
with ${\leadsto_\nt}$.

\begin{theorem}[Proving Non-Termination via ${\leadsto_\nt}$]
  \label{thm:nonterm}
  Let $f(\vec{x}) \xto{} f(\vec{a}) \cond{\phi} \in \TT$. 
  If $\prob{\top}{\top}{\phi}{\vec{a}} \leadsto_\nt^* \prob{\psi}{\phi}{\top}{\vec{a}}$,
  then for every $\vec{c} \in \ZZ^d$ where $\psi(\vec{c})$ is satisfiable, the
  configuration $f(\vec{c})$ admits\report{
 a $\xto{}_{\TT}$-sequence where the transition $f(\vec{x}) \xto{} f(\vec{a}) \cond{\phi}$
 is applied infinitely often}\submission{ an
         infinite $\xto{}_{\TT}$-sequence}.
\end{theorem}
\report{
  \begin{proof}
    If $\TV(\phi) \cup \TV(\vec{a}) = \emptyset$, then the claim is equivalent to \cite[Thm.\ 14]{sttt21}.
    If $\TV(\phi) \cup \TV(\vec{a}) \neq \emptyset$, then the claim follows by considering the variables $\TV(\phi) \cup \TV(\vec{a})$ as additional program variables, whose update is the identity.
    \qed
  \end{proof}
}

The conditional non-termination techniques used by \tool{LoAT} are {\bf Increase}, {\bf Eventual Increase}, and {\bf Fixpoint}.
So non-termination proofs can be synthesized while trying to accelerate a loop with very little overhead.
After successfully accelerating a loop as explained in \Cref{sec:sat}, \tool{LoAT} tries to find a second suitably ordered ${\leadsto}$-sequence, where it only considers the conditional non-termination techniques mentioned above.
If \tool{LoAT} succeeds, then it has found a ${\leadsto_\nt}$-sequence\report{
  $\prob{\top}{\top}{\phi}{\vec{a}} \leadsto_\nt^* \prob{\psi}{\phi}{\top}{\vec{a}}$.
  If $\psi$ is satisfiable, then \tool{LoAT} has successfully proven non-termination of the loop under consideration due to \Cref{thm:nonterm}.
}\submission{
  which gives rise to a proof of non-termination via \Cref{thm:nonterm}.
}

\report{
  \begin{theorem}[Soundness of Non-Termination Processor]
    \label{thm:nonterm-proc}
    If $f(\vec{x}) \xto{p} f(\vec{a}) \cond{\phi} \in \TT$, $\prob{\top}{\top}{\phi}{\vec{a}} \leadsto_\nt^* \prob{\psi}{\phi}{\top}{\vec{a}}$, and ${} \models \phi \implies p > 0$ then the processor which transforms $\TT$ to
    \[
      \TT \cup \{f(\vec{x}) \xto{\infty} \sink(\vec{x}) \cond{\psi}\}
    \]
    is sound for lower bounds and non-termination.
  \end{theorem}
  \report{
    \begin{proof}
      To show the soundness for lower bounds, note that $\dt_\TT(f'(\vec{c}\,'))$
      and\linebreak
      $\dt_{\TT \cup \{f(\vec{x}) \xto{\infty} \sink(\vec{x}) \cond{\psi}\}}(f'(\vec{c}\,'))$ only differ for configurations $f'(\vec{c}\,')$ which can reach a configuration $f(\vec{c})$ in $\TT$ where $\psi(\vec{c})$ is satisfiable.
      Then by \Cref{thm:nonterm}, $f'(\vec{c}\,')$ admits an infinite
      ${\xto{}_\TT}$-sequence where
        a transition with cost $p$ and guard $\phi$ is applied infinitely many
        times.        
      By $\models \phi \implies p > 0$, the cost of this infinite sequence is $\infty$ and thus, we have $\dt_\TT(f'(\vec{c}\,')) = \infty$.

      The processor is also sound for non-termination.
If  $\TT$ is finitary but   $\TT \cup \{f(\vec{x}) \xto{\infty} \sink(\vec{x}) \cond{\psi}\}$ is not,
then there is a configuration $\start(\vec{c}\,')$ which
reaches a configuration $f(\vec{c})$ in $\TT$ where $\psi(\vec{c})$ is satisfiable. Again,
by
\Cref{thm:nonterm}, then $f(\vec{c})$ and hence also
$\start(\vec{c}\,')$
admit an infinite ${\xto{}_\TT}$-sequence where a
transition with cost $p$ and guard $\phi$ is applied infinitely many times.
 But by $\models \phi \implies p > 0$, the cost of this infinite sequence is $\infty$
 which contradicts the finitism of $\TT$.
      \qed
    \end{proof}
 }

  Note that \tool{LoAT} cannot detect non-terminating loops with non-positive costs, since it may only apply \Cref{thm:nonterm-proc} if ${} \models \phi \implies p > 0$.
  Without this precondition, the processor from \Cref{thm:nonterm-proc} would not be sound for lower bounds.

  \begin{example}
    \label{nontermLargeExample}
    We show how \tool{LoAT} proves non-termination of \cite[Ex.\ 17]{sttt21}:
    \[
      f(x_1,x_2,x_3,x_4) \to f(1,x_2 + x_1,x_3 + x_2,-x_4) \cond{x_1 > 0 \land x_3 > 0 \land x_4 + 1 > 0}
    \]
    We have $\encode_{\inc,x_1>0} = (x_1 > 0 \land x_3 > 0 \land x_4 + 1 > 0 \implies 1 > 0)$, which is valid.
    Assuming that the SMT solver finds the unsat core $\{1 \leq 0\}$ of $\neg\encode_{\inc,x_1>0}$, we get $\deps_{\inc,x_1>0} = \emptyset$, such that \tool{LoAT} does not try to apply any other acceleration techniques to $x_1 > 0$.

    Moreover, we have
    \begin{align*}
      \encode_{\inc,x_3>0} = {}   & (x_1 > 0 \land x_3 > 0 \land x_4 + 1 > 0 \implies x_3 + x_2 > 0),                                                  \\
      \encode_{\dec,x_3>0} = {}   & (x_1 > 0 \land x_3 + x_2 > 0 \land x_4 + 1 > 0 \implies x_3 > 0),                                     \; \text{and} \\
      \encode_{\evdec,x_3>0} = {} & (x_3 \geq x_3 + x_2 \land x_1 > 0 \land x_4 + 1 > 0 \implies x_3 + x_2 \geq x_3 + 2 \cdot x_2 + x_1),
    \end{align*}
    which are all invalid.
    However,
    \begin{align*}
                   & \encode_{\evinc,x_3>0}                                                                               \\
      {} = {}      & (x_3 \leq x_3 + x_2 \land x_1 > 0 \land x_4 + 1 > 0 \implies x_3 + x_2 \leq x_3 + 2 \cdot x_2 + x_1) \\
      {} \equiv {} & (x_2 \geq 0 \land x_1 > 0 \land x_4 + 1 > 0 \implies x_2 + x_1 \geq 0)
    \end{align*}
    is valid.
    Assuming that the SMT solver finds the unsat core
    \[
      \{x_2 \geq 0,x_1 > 0,x_2 + x_1 < 0\}
    \]
    of $\neg \encode_{\evinc,x_3>0}$, we get $\deps_{\evinc,x_3>0} = \{x_1>0\}$.
    Thus, \tool{LoAT} does not try to apply {\bf Fixpoint} to $x_3 > 0$.

    Finally, all of
    \begin{align*}
      \encode_{\inc,x_4+1>0} = {}   & (x_1 > 0 \land x_3 > 0 \land x_4 + 1 > 0 \implies -x_4 + 1 > 0),                 \\
      \encode_{\dec,x_4+1>0} = {}   & (x_1 > 0 \land x_3 > 0 \land -x_4 + 1 > 0 \implies x_4 + 1 > 0),                 \\
      \encode_{\evdec,x_4+1>0} = {} & (x_4 \geq -x_4 \land x_1 > 0 \land x_3 > 0 \implies -x_4 \geq x_4), & \text{and} \\
      \encode_{\evinc,x_4+1>0} = {} & (x_4 \leq -x_4 \land x_1 > 0 \land x_3 > 0 \implies -x_4 \leq x_4)
    \end{align*}
    are invalid.
    However,
    \[
      \accel_\fixpoint(x_4+1>0, x_1 > 0 \land x_3 > 0, \vec{a}) = (x_4+1>0 \land x_4 = -x_4) \equiv (x_4 = 0)
    \]
    is satisfiable.
    This suffices to conclude
    \[
      \{(\inc,x_1>0),(\evinc,x_3>0),(\fixpoint,x_4+1>0)\} = S \subseteq C
    \]
    and $(\inc,x_1>0) \prec (\evinc,x_3>0)$, see \Cref{lem:order}.
    By choosing, e.g., the strict total order with $(\inc,x_1>0) \prec' (\evinc,x_3>0) \prec' (\fixpoint,x_4+1>0)$ containing ${\prec}$, we get the ${\leadsto_\nt}$-sequence
    \begin{align*}
      \prob{\top}{\top}{\phi}{\vec{a}} \leadsto_\nt {} & \prob{x_1>0}{x_1>0}{x_3>0 \land x_4+1>0}{\vec{a}}                              \\
      \leadsto_\nt {}                                  & \prob{x_1>0 \land x_3>0 \land x_2 \geq 0}{x_1>0 \land x_3>0}{x_4+1>0}{\vec{a}} \\
      \leadsto_\nt {}                                  & \prob{x_1>0 \land x_3>0 \land x_2 \geq 0 \land x_4 = 0}{\phi}{\top}{\vec{a}}
    \end{align*}
    due to \Cref{thm:order}.
    Since $x_1>0 \land x_3>0 \land x_2 \geq 0 \land x_4=0$ is satisfiable, \tool{LoAT} accelerates the original transition to
    \[
      f(x_1,x_2,x_3,x_4) \xto{\infty} \sink(x_1,x_2,x_3,x_4) \cond{x_1>0 \land x_3>0 \land x_2 \geq 0 \land x_4=0}.
    \]
  \end{example}
}

\submission{\section{Implementation, Experiments and Conclusion}}\report{\section{Implementation and Experiments}}
\label{sec:experiments}

Our implementation in \tool{LoAT} can parse three widely used formats for ITSs (see \cite{website}), and it is configurable via a minimalistic set of command-line options:
\begin{description}
\item[{\tt --timeout}] to set a timeout in seconds
\item[{\tt --proof-level}] to set the verbosity of the proof output
\item[{\tt --plain}] to switch from colored to monochrome proof-output
\item[{\tt --limit-strategy}] to choose a strategy for solving limit problems, see \cite{loat-journal}
\item[{\tt --mode}] to choose an analysis mode for \tool{LoAT} ({\tt complexity} or {\tt non\_termination})
\end{description}
We evaluate three versions of \tool{LoAT}:
\tool{LoAT~'19} uses templates to find invariants that facilitate loop acceleration for proving non-termination \cite{fmcad19};
\tool{LoAT~'20} deduces worst-case lower bounds based on loop acceleration via
\emph{metering functions} \cite{loat-journal};
and \tool{LoAT~'22}  applies the calculus from \cite{tacas20,sttt21} as described
in \Cref{sec:sat,sec:nonterm}.
We also include three other state-of-the-art termination tools in our evaluation: \tool{T2} \cite{t2-tool}, \tool{VeryMax} \cite{larraz14}, and \tool{iRankFinder}
\cite{amram18,irankWST}.
Regarding complexity, the only other tool for worst-case lower bounds of ITSs is \tool{LOBER} \cite{Albert21}.
However, we do not compare with \tool{LOBER}, as it only analyses (multi-path) loops instead of full ITSs.
\report{We do not include \tool{AProVE} \cite{tool-jar}, since it can only prove non-termination of integer transition systems without dedicated start locations, and hence a comparison would not be meaningful.
  We also exclude \tool{RevTerm} \cite{Chatterjee21}, since the evaluation from
  \cite{sttt21} showed that it is not competitive in a sequential setting.
  Finally, we also exclude \tool{Ultimate} \cite{Ultimate}, since it analyzes {\sf C} programs, whereas \tool{LoAT} analyzes integer transition systems and thus \tool{Ultimate} and \tool{LoAT} do not have a common input format.
}

We use the examples from the categories \emph{Termination} (1222 examples) and \emph{Complexity of ITSs} (781 examples), respectively, of the \emph{Termination Problems Data Base} \cite{tpdb}\report{, the collection of benchmarks that is used in the annual \emph{Termination and Complexity Competition} (\emph{TermComp}) \cite{termcomp}}. 
All benchmarks have been performed on \emph{StarExec} \cite{starexec} (Intel Xeon E5-2609, 2.40GHz, 264GB RAM \cite{starexec-spec}) with a wall clock timeout of
300~s.
\report{For \tool{T2} and \tool{VeryMax}, we took the versions of their last \emph{TermComp} participations (2015 and 2019, respectively).
  We used a configuration of \tool{iRankFinder} that is tailored towards proving non-termination after consulting its authors.}

\smallskip
\noindent
\hfill
\begin{minipage}{0.5\textwidth}
  \begin{adjustbox}{width=\textwidth,center}
    \begin{tabular}{|c||c|c|c|c|c|}
      \hline                        & No  & Yes & Avg.\ Rt & Median Rt & Std.\ Dev.\ Rt \\
      \hline \hline \tool{LoAT~'22} & 493 & 0   & 9.4      & 0.2 & 41.5 \\
      \hline \tool{LoAT~'19}        & 459 & 0   & 22.6     & 1.5 & 67.5 \\
      \hline \tool{T2}              & 438 & 610 & 22.6     & 1.2 & 66.7 \\
      \hline \tool{VeryMax}         & 419 & 628 & 29.9     & 1.0 & 66.7 \\
      \hline \tool{iRankFinder}     & 399 & 634 & 44.1     & 4.9 & 89.1 \\
      \hline
    \end{tabular}
  \end{adjustbox}
\end{minipage}
\hfill
\begin{minipage}{0.45\textwidth}
  \newcommand{\Om}{\Omega}
  \begin{adjustbox}{width=\textwidth,center}
    \begin{tabular}{|l||l|c|c|c|c|c|c|}
      \cline{2-8}
      \multicolumn{1}{c}{} & \multicolumn{7}{|c|}{\tool{LoAT~'22}}                                                                            \\
      \hhline{-=======}
      \multirow{8}{*}{\rotatebox[origin=c]{90}{\pbox[c]{100em}{
            \centering
            \tool{LoAT~'20}}}}
                           & $\rc_{\TT}(n)$                        & $\Om(1)$ & $\Om(n)$ & $\Om(n^2)$ & $\Om(n^{>2})$ & $EXP$ & $\Om(\omega)$ \\
      \cline{2-8}
                           & $\Om(1)$                              & $180$    & $63$     & $1$        & $-$           & $-$   & $12$          \\
      \cline{2-8}
                           & $\Om(n)$                              & $6$      & $218$    & $3$        & $-$           & $-$   & $-$           \\
      \cline{2-8}
                           & $\Om(n^2)$                            & $-$      & $1$      & $69$       & $-$           & $-$   & $-$           \\
      \cline{2-8}
                           & $\Om(n^{>2})$                         & $-$      & $-$      & $-$        & $7$           & $-$   & $-$           \\
      \cline{2-8}
                           & $EXP$                                 & $1$      & $-$      & $-$        & $-$           & $4$   & $-$           \\
      \cline{2-8}
                           & $\Om(\omega)$                         & $-$      & $-$      & $-$        & $-$           & $-$   & $216$         \\
      \hline
    \end{tabular}
  \end{adjustbox}
\end{minipage}
\hfill
\smallskip

By the table on the left, \tool{LoAT~'22} is the most powerful tool for non-termination.
The improvement over \tool{LoAT~'19} demonstrates that the calculus
from \cite{tacas20,sttt21} is more powerful and efficient than the approach
from \cite{fmcad19}.
The last three columns show the average, the median, and the standard deviation of the wall clock runtime, including examples where the timeout was reached.
\report{In addition, there are further minor aspects in which
\tool{LoAT~'22} improves over its predecessors.}
\report{The most important one is the new SMT interface in \tool{LoAT~'22}, which allows for the integration of several solvers, such that \tool{LoAT} can now choose a suitable solver for each SMT query.
  Currently, \tool{LoAT} uses \tool{Yices} \cite{yices}
  for linear arithmetic and \tool{Z3} \cite{z3} for non-linear arithmetic.}

The table on the right shows the results for complexity.
The diagonal corresponds to examples where \tool{LoAT~'20} and \tool{LoAT~'22} yield the
same result. The entries above or below the diagonal correspond to examples where \tool{LoAT~'22} or \tool{LoAT~'20} is better, respectively.
\report{There are $8$ regressions (i.e., entries below the diagonal), which are due to various reasons.
  For most of them, the more powerful acceleration techniques of \tool{LoAT~'22} lead to more transitions during \tool{LoAT}'s program simplification, such that the ITS gets too large and hence \tool{LoAT} runs out of resources.
  In other cases, \tool{LoAT~'22} fails since some heuristics with little positive impact have been removed to simplify the code base.
However, in $79$ cases \tool{LoAT~'22} yields better results than \tool{LoAT~'20}, i.e.,
  the integration of the calculus from \cite{tacas20,sttt21} in \tool{LoAT~'22} is also very beneficial for
  deducing worst-case lower bounds.    For further details on our evaluation, we refer
  to \cite{website}.}%
\submission{There are $8$ regressions and $79$ improvements, so the calculus from
  \cite{tacas20,sttt21} used by \tool{LoAT~'22} is also beneficial for lower bounds.} 

\report{Regarding the runtime of the tool, \tool{LoAT~'20} is considerably
  faster than \tool{LoAT~'22} (6.0~s vs.\ 15.5~s on average).
  The reason is that \tool{LoAT~'20} accelerates loops via metering functions
  \cite{loat-journal},
  a technique that can be implemented with a single SMT query.
  In contrast, the calculus from \cite{tacas20,sttt21} requires several SMT queries.
  So \tool{LoAT~'22} is faster than \tool{LoAT~'19}, but slower than \tool{LoAT~'20}, and substantially more powerful than both \tool{LoAT~'19} and \tool{LoAT~'20}.

  As a sanity check, we also compared the results of \tool{LoAT~'22} with the worst-case
  \emph{upper} bounds deduced by a current version of the tool \tool{KoAT} \cite{koat,Festschrift}.
  There were no conflicts, i.e., there is no example where the lower bound inferred
  by \tool{LoAT} exceeds the upper bound proven by \tool{KoAT}.
}

\submission{
  \tool{LoAT} is open source and its source code is available on GitHub \cite{loat-github}.
  See \cite{website,report} for details on our evaluation, related work, all proofs, and a pre-compiled binary.
}

\report{
  \section{Related Work}
  \label{sec:related}

  Loop acceleration has successfully been used to prove various properties like reachability, safety, (non-)termination, and lower runtime bounds in the past\cite{loat-journal,tacas20,sttt21,underapprox15,bozga10,bozga09a}.
  See \cite{sttt21} for a detailed comparison of existing loop acceleration techniques and techniques for proving non-termination, including the calculus from \cite{tacas20,sttt21} that is used by \tool{LoAT}.
  The main novelty of our approach in the current paper is that we integrated this calculus into a whole-program-analysis framework for the first time, and we presented a novel technique based on SMT solving and unsat cores to implement it efficiently.

  Regarding the competing tools for proving non-termination from \Cref{sec:experiments}, the main differences to \tool{iRankFinder}, \tool{T2}, and \tool{VeryMax} are as follows.
  To find witnesses of non-termination for loops, \tool{iRankFinder} iteratively proves termination of configurations by deducing ranking functions until termination of all configurations has been proven or a sufficient condition for non-termination is found (or the analysis diverges).
  \tool{T2} and \tool{VeryMax} use template-based approaches.
  In contrast, \tool{LoAT} uses the calculus from \cite{sttt21}.
  For proving reachability of these witnesses of non-termination, both \tool{iRankFinder} and \tool{VeryMax} use off-the-shelf tools and \tool{T2} uses reverse evaluation.
  In contrast, \tool{LoAT} uses the approach from \cite{fmcad19,loat-journal} based on loop acceleration.

  Regarding
  lower bounds, note that the \emph{worst-case} lower bounds computed by \tool{LoAT} differ fundamentally from the \emph{best-case} lower bounds computed by, e.g., \tool{PUBS} \cite{pubs-upper-lower} and \tool{CoFloCo} \cite{cofloco2}:
  Best-case lower bounds are valid \emph{for all} program runs and hence they are interesting, e.g., in the context of scheduling (see \cite{pubs-upper-lower}).
  Worst-case lower bounds are valid for \emph{some} (usually infinitely many) program runs and hence they are useful for finding performance bugs.

  To the best of our knowledge, the only other approach for deducing worst-case lower bounds for programs operating on integers is \cite{Albert21}.
  However, this approach focuses on stand-alone loops, whereas \tool{LoAT} analyzes whole programs.
  Moreover, the approach from \cite{Albert21} is tailored towards a precise handling of multi-path loops, whereas our techniques for dealing with loops from \cite{tacas20,sttt21}
  focus on single-path loops (see \cite{fmcad19,loat-journal} for \tool{LoAT}'s strategy to
  analyze full programs, including multi-path loops, with the processors from \Cref{sec:overview}).
  Integrating the approach from \cite{Albert21} into an under-approximating whole-program analysis framework like ours is challenging, since it is usually not possible to compute closed forms for multi-path loops.

 In \cite{jar17}, we presented techniques to deduce worst-case lower bounds for \emph{term rewrite systems}, which are orthogonal to the ITSs considered in the current paper:
  They allow for full recursion and tree-shaped data structures, but they do not support integer arithmetic natively.

  Apart from that, the most closely related techniques to our approach for proving lower bounds are techniques for proving non-termination.
  Here, the crucial difference is that a single witness suffices to prove non-termination, whereas an infinite set of witnesses is required to prove a non-constant lower bound for a terminating ITS.
}

\report{
  \section{Conclusion}
\label{sec:conclusion}

We presented the newest version of the \emph{Loop Acceleration Tool} \tool{LoAT}.
Currently, \tool{LoAT} is the most powerful tool for proving non-termination of programs operating on integers, as witnessed by our evaluation (\Cref{sec:experiments}) as well as the annual \emph{Termination and Complexity Competition} \cite{termcomp}.
Consequently, it is also used in the backend for proving non-termination of {\sf C} programs
by \tool{AProVE} \cite{SVCOMP22}.
Moreover, it is the only tool for deducing worst-case lower bounds for general programs operating on integers.
\tool{LoAT} is open source, its source code is available on GitHub,
\begin{center}
  \url{https://github.com/aprove-developers/LoAT},
\end{center}
and
we refer to \cite{website} for
a pre-compiled binary of \tool{LoAT}.
 
In future work, we plan to extend \tool{LoAT}'s under-approximations to programs written in (a fragment
of) the programming language {\sf C}.
}

\bibliographystyle{splncs04}
\submission{\bibliography{refs,crossrefs,strings}}

\report{\providecommand{\noopsort}[1]{}

}

\end{document}